\documentclass[aps,prb,twocolumn,groupedaddress,showpacs]{revtex4-1}
\Large   
\usepackage{graphicx}

\begin{document}

\title{Characterization of the Heavy Metal Pyrochlore Lattice Superconductor CaIr$_{2}$}

\author{Neel Haldolaarachchige, Quinn Gibson, Leslie M.~Schoop, Huixia Luo and R.~J.~Cava}
\address{Department of Chemistry, Princeton University, Princeton, New Jersey 08544, USA}

\begin{abstract}
We report the electronic properties of the cubic laves phase superconductor CaIr$_{2}$ ($T_{c}=5.8$~K), in which the Ir atoms have a Pyrochlore lattice. The estimated superconducting parameters obtained from magnetization and specific heat measurements indicate that CaIr$_{2}$ is a weakly coupled BCS superconductor. Electronic band structure calculations show that the Ir \textit{d}-states are dominant at the Fermi level, creating a complex Fermi surface that is impacted substantially by spin orbit coupling. 

\end{abstract}

\maketitle
\newcommand{\angstrom}{\mbox{\normalfont\AA[12pt,a4paper,final]}}

\section{Introduction}

Compounds based on 5\textit{d} transition metals are of recent interest because electron correlations and spin-orbit interactions play an important role in determining their electronic properties. Iridium oxides with the pyrochlore lattice, in particular, are predicted to host exotic electronic states~\cite{iridate}, but these materials have not yet been shown to host superconductivity.  
A handful of Ir compounds are known to be superconducting, some more likely showing this property due to the presence of rare earths rather than the Ir, but in other cases, such as for IrGe and Mg$_{10}$Ir$_{19}$B$_{16}$, the superconductivity is derived from Ir states at the Fermi Energy~\cite{ceirin5,daigo2013,ca3irge32014,cairge32010,daigo2014,mgirb}. 

Here we report the synthesis, experimental electronic characterization, and calculated electronic band structure of the cubic Laves phase superconductor CaIr$_{2}$. The Ca atoms in this material (see Fig.~\ref{Fig1}(b)) occupy the positions of the diamond structure and the Ir atoms (see Fig.~\ref{Fig1}(c)) form a pyrochlore lattice. The Ir-Ir separation in the pyrochlore network is 2.67 \AA, which is smaller than that in elemental Ir (2.76 \AA)~\cite{elis1958}.
The existence of superconductivity in CaIr$_{2}$ has been reported earlier, but only its T$_{c}$ - no further characterization is available.~\cite{mattias1957} 
The reported T$_c$ of CaIr$_{2}$ is relatively high for an Ir-based compound, at 5.8 K. CeRu$_{2}$, a cubic Laves phase that we employ for comparison purposes here, displays unusual superconducting properties due to the presence of Ce 4\textit{f}  states at the Fermi Energy.~\cite{ceru2-1, ceru2-2}

\section{Experiment and Calculation}
Polycrystalline samples of CaIr$_{2}$ were prepared by a two-step solid state reaction method, starting from elemental Ca-pieces (99.99$\%$; Alfa Aesar) and Ir-powder (99.99$\%$; Alfa Aesar). A significant amount of calcium excess (200$\%$) was found to be necessary to make high quality material due to Ca volatilization. The starting materials were added into an alumina crucible inside an Ar-filled glove box, which was then sealed inside an evacuated quartz tube. The tube was then slowly heated (50~$^{0}$C per hour) and held at 850~$^{0}$C for 20 hrs. The resulting powder was then ground well and pressed into a pellet, again with excess calcium, slowly heated in an evacuated quartz tube to 700~$^{0}$C, and then held for 12 hrs. The material was single phase at this point, and was kept inside the glove box until characterization. Such handling is necessary to avoid decomposition.\cite{satya2014, neel2014-2} The purity and cell parameters of the samples were evaluated by powder X-ray diffraction (PXRD) at room temperature on a Bruker D8 FOCUS diffractometer (Cu$~K_{\alpha}$). 

The electrical resistivity was measured using a standard four-probe method with an excitation current of 10 mA; small diameter Pt wires were attached to the sample using a conductive epoxy (Epotek H20E). Data were collected from 300 - 2 K and in magnetic fields up to 5 T using a Quantum Design Physical Property Measurement System (PPMS). The specific heat was measured between 2 and 20 K in the PPMS, using a time-relaxation method, at 0 and 5 Tesla applied magnetic fields. The magnetic susceptibility was measured in a DC
field of 10 and 100 Oe; the sample was cooled down to 2 K in zero-field, the magnetic field was then applied, and the sample magnetization followed on heating to 8 K [zero-field-cooled (ZFC)], and then on cooling to 2 K [field-cooled (FC)] in the PPMS.
The electronic structure calculations were performed by density functional theory (DFT) using the WIEN2K code with a full-potential linearized augmented plane-wave and local orbitals [FP-LAPW + lo] basis~\cite{blaha2001,sign1996,madsen2001,sjosted2000} together with the PBE parameterization~\cite{perdew1996} of the GGA, with and without spin orbit coupling (SOC). The plane-wave cutoff parameter R$_{MT}$K$_{MAX}$ was set to 7 and the Brillouin zone was sampled by 20,000 k points. 
The Fermi surface was plotted with the program Crysden.

\section{Results and Discussion}
\begin{figure}[t]
  \centerline{\includegraphics[width=0.5\textwidth]{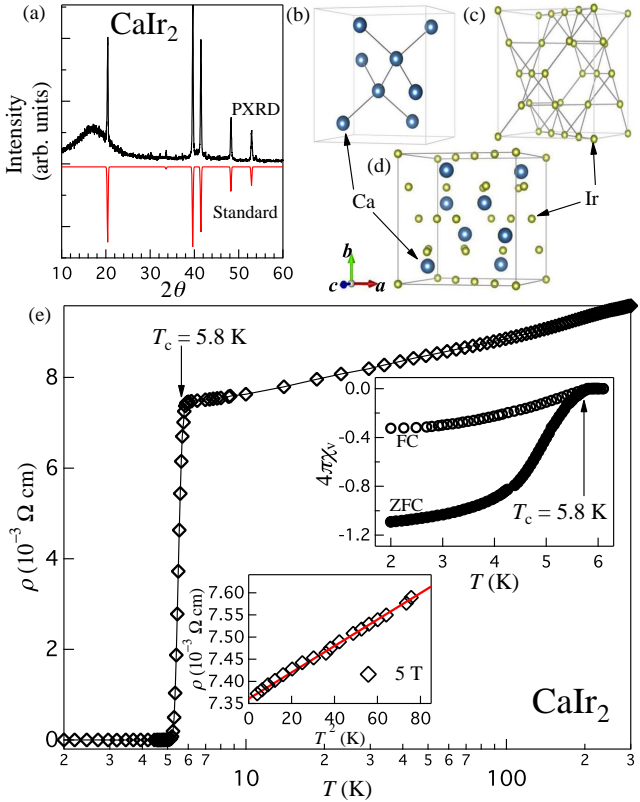}}
  \caption
    {
      (Color online) Structure and Elementary Characterization of CaIr$_{2}$.  (a) the PXRD pattern of CaIr$_{2}$. (b) Represents the diamond structure for the Ca atoms CaIr$_{2}$, while (c) shows the corner sharing tetrahedral network of of Ir atoms and (d) shows the full 3D crystal structure. (e) Semi log plot of resistivity as a function of temperature for CaIr$_{2}$. The upper right inset shows the DC-magnetization (ZFC and FC) as a function of temperature. The lower inset shows the resistivity as a function of \textit{T}$^{2}$. The solid line in lower left inset represents a linear fit to the data.
    }
  \label{Fig1}
\end{figure}

Fig.~\ref{Fig1} shows the PXRD analysis of the polycrystalline sample of CaIr$_{2}$ employed for the characterization, the 3D crystal structure, the resistivity, and the DC-magnetization.  Fig.~\ref{Fig1}(a) shows that the PXRD pattern of the CaIr$_{2}$ sample is an excellent match to the standard pattern in the ICSD database (code number 108146). (The 'hump' in the low 2$\theta$ range of the PXRD pattern is due to the paratone-oil that covers the sample to prevent it from decomposing during the acquisition of the pattern.) A schematic view of the crystal structure of CaIr$_{2}$ is shown in Fig.~\ref{Fig1}(d) The pyrochlore lattice of Ir atoms is emphasized.

\begin{figure}[t]
  \centerline{\includegraphics[width=0.5\textwidth]{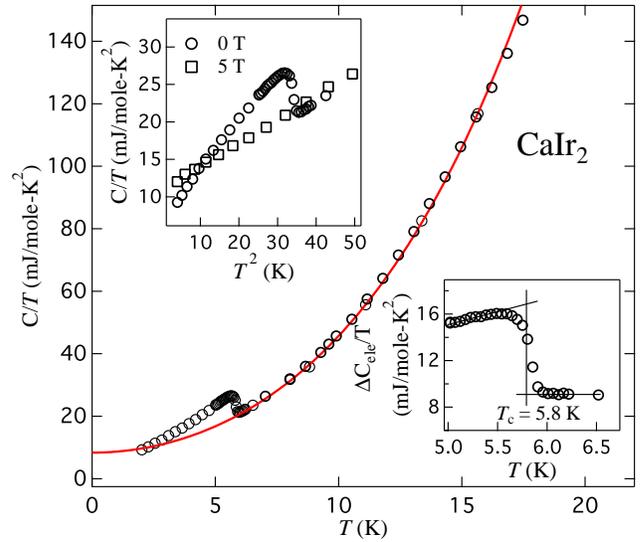}}
  \caption
    {
      (Color online)  Analysis of the heat capacity of CaIr$_{2}$. (a) The main panel shows the heat capacity in 0 T applied field. The solid line shows the heat capacity data fit with the equation $C=\gamma T+\beta_{1} T^{3}+\beta_{2} T^{5}$. The upper left inset shows the heat capacity as a function of \textit{T}$^{2}$ at zero applied field and 5 T applied field. The lower right inset shows the electronic heat capacity jump at T$_{c}$. 
    }
  \label{Fig2}
\end{figure}

The main panel of Fig.~\ref{Fig1}(e) shows the temperature dependent resistivity of CaIr$_{2}$ from 300 to 2 K. Poor metallic behavior $\left( \frac{d\rho}{dT} > 0\right)$ can be observed in the normal state resistance of the polycrystalline sample, with a room temperature $\rho \sim 9.6~m\Omega~cm$. A clear superconducting transition can be observed at 5.8 K.
The low-temperature resistivity data (see lower left inset of the Fig.~\ref{Fig1}(e)) can be described by a power law $\rho = \rho _{0} + AT^{n}$ with $n=2$, which follows Fermi liquid behavior.
The upper right inset of the Fig.~\ref{Fig1}(e) shows the DC-magnetization of CaIr$_{2}$. The superconducting shielding can be observed in the zero-field-cooled (ZFC-shielding) and field-cooled (FC-Meissner) data in the figure. The bulk superconducting transition T$^{onset} _{c}$ = 5.8 K can clearly be observed. The very similar values of T$_{c}$ seen in both resistivity and susceptibility indicates that the polycrystalline sample is nearly homogeneous.~\cite{tinkham}

Fig.~\ref{Fig2} shows the characterization of the superconducting transition by specific heat measurements. The main panel of the Fig.~\ref{Fig2} shows $\frac {C}{T}$ as a function of $T$, characterizing the specific heat jump at the thermodynamic transition. This jump is completely suppressed under a 5 T applied magnetic field. The superconducting transition temperature $T_{c}$ = 5.8 K is shown in the lower right inset of Fig.~\ref{Fig2}, as extracted by the standard equal area construction method. 
The low temperature normal state specific heat is non-Debye-like. Non-Debye behavior has often been reported in superconductors and has been ascribed either to a large Einstein contribution or a low Debye temperature $\theta _{D}$. Because an Einstein phonon contribution is negligible below 20 K, a second term in the harmonic-lattice approximation is needed to improve the fit to the specific heat data. Similar behavior has been observed on some other heavy element superconductors such as BaBi$_{3}$.~\cite{neel2014-2}
We find that the low temperature normal state specific heat can be well fitted with $\frac{C}{T} = \gamma _{n} + \beta_{1} T^{2} + \beta_{2} T^{4}$,
where $\gamma _{n} T$ represents the electronic contribution in the normal state and $\beta_{1} T^{3}$ and $\beta_{2} T^{5}$ describe the lattice-phonon contributions to the specific heat. 
The solid line in Fig.~\ref{Fig2} shows the fitting; the electronic specific heat coefficient $\gamma _{n} = 8.36 \frac{mJ}{mol~K^{2}}$ and the phonon/lattice contributions $\beta_{1} = 0.32 \frac{mJ}{mol~K^{4}}$ and $\beta_{2} = 0.00048 \frac{mJ}{mol~K^{4}}$ are extracted from the fit. 
The value of $\gamma$ obtained is smaller than that of cubic Laves phase superconductor CeRu$_{2}$, (which may be due to the fact that \textit{f} orbitals are not present in the current case) but is comparable to some other alkaline/iridium based heavy element superconductors (see Table.~\ref{tab:1}).~\cite{ceru2-2, satya2014, daigo2014} 

\begin{figure}[t]
  \centerline{\includegraphics[width=0.5\textwidth]{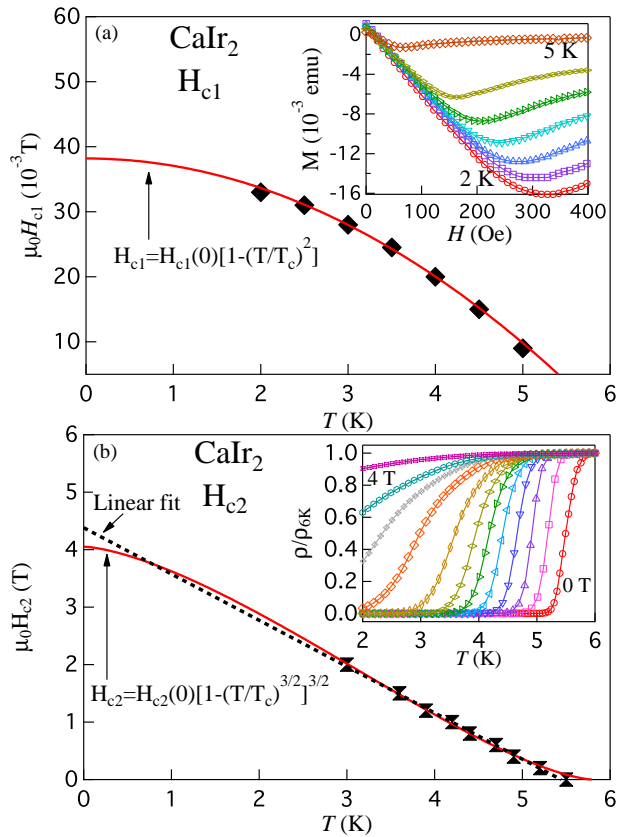}}
  \caption
    {
      (Color online) Analysis of lower and upper critical fields of the cubic Laves phase superconductor CaIr$_{2}$. (a) The lower critical field as a function of temperature. The inset shows the DC-magnetization as a function of applied magnetic field at different temperatures below the superconducting T$_{c}$. (b) The upper critical field as a function of temperature. Inset shows the resistivity with increasing applied magnetic field.
    }
  \label{Fig3}
\end{figure}

Fig.~\ref{Fig3} shows the analysis of the lower and upper critical fields of CaIr$_{2}$. The magnetization as a function of magnetic field over a range of temperatures below the superconducting T$_{c}$ is shown in the upper right inset of Fig.~\ref{Fig3}(a). For analysis of the lower critical field, the point of 2.5\% deviation from the full shielding effect was selected for each temperature. The main panel of the Fig.~\ref{Fig3}(a) shows $\mu_{0}H_{c1}$ as a function of T$_{c}$. The lower critical field behavior was analyzed with the equation $H_{c1}(T)=H_{c1}(0)\left[1-\left(\frac{T}{T_{c}}\right)^{2}\right]$. The $\mu _{0}H_{c1}$ data is well described with the above equation, and a least squares fit yielded the value of $\mu _{0}H_{c1}(0)$=381 Oe, which is comparable with the cubic Laves phase CeRu$_{2}$ (see Table.~\ref{tab:1}).  

Fig.~\ref{Fig3}(b) shows the magnetoresistance data for CaIr$_{2}$. The width of the superconducting transition increases slightly with increasing magnetic field. Selecting the 50$\%$ normal state resistivity point as the transition temperature, we estimate the orbital upper critical field, $\mu _{0}H_{c2}$(0), from the Werthamer-Helfand-Hohenberg (WHH) expression, 
$\mu _{0}H_{c2}(0)=-0.693~T_{c}\frac{dH_{c2}}{dT}\vert_{T=T_{c}}$. A nearly linear relationship is observed in Fig.~\ref{Fig3}(b) between $\mu _{0}H_{c2}$ and $T_{c}$. The slope is used to calculate $\mu _{0}H_{c2}(0)=$~4.0~T. This value of $\mu _{0}H_{c2}(0)$ is smaller than the weak coupling Pauli paramagnetic limit $\mu _{0}H^{Pauli} = 1.82~T_{c} = 10.6$ T for CaIr$_{2}$. 
We also used the empirical formula $H_{c2}(T)=H_{c2}(0)\left[1-\left(\frac{T}{T_{c}}\right)^{\frac{3}{2}}\right]^{\frac{3}{2}}$
to calculate orbital upper critical field $\left( \mu _{0}H_{c2}(0)=3.8 T\right) $, which yields a value that agrees well with the calculated value using the WHH method. The WHH expression and the empirical formula are widely used to calculate the $\mu _{0}H_{c2}(0)$ for a variety of intermetallic and oxide superconductors.~\cite{amar2011, lan2001, neel2014, daigo2013} 

The upper critical field value $\mu _{0}H_{c2}(0)$ can be used to estimate the Ginzburg-Landau coherence length $\xi (0)=\sqrt{\Phi _{0}/2\pi H_{c2}(0)}=99$~\AA, where $\Phi _{0}=\frac{hc}{2e}$ is the magnetic flux quantum.~\cite{clogston1962, werthamer1966} This value is higher than that of the Laves phase CeRu$_{2}$ (see Table.~\ref{tab:1}).

\begin{figure}[t]
  \centerline{\includegraphics[width=0.5\textwidth]{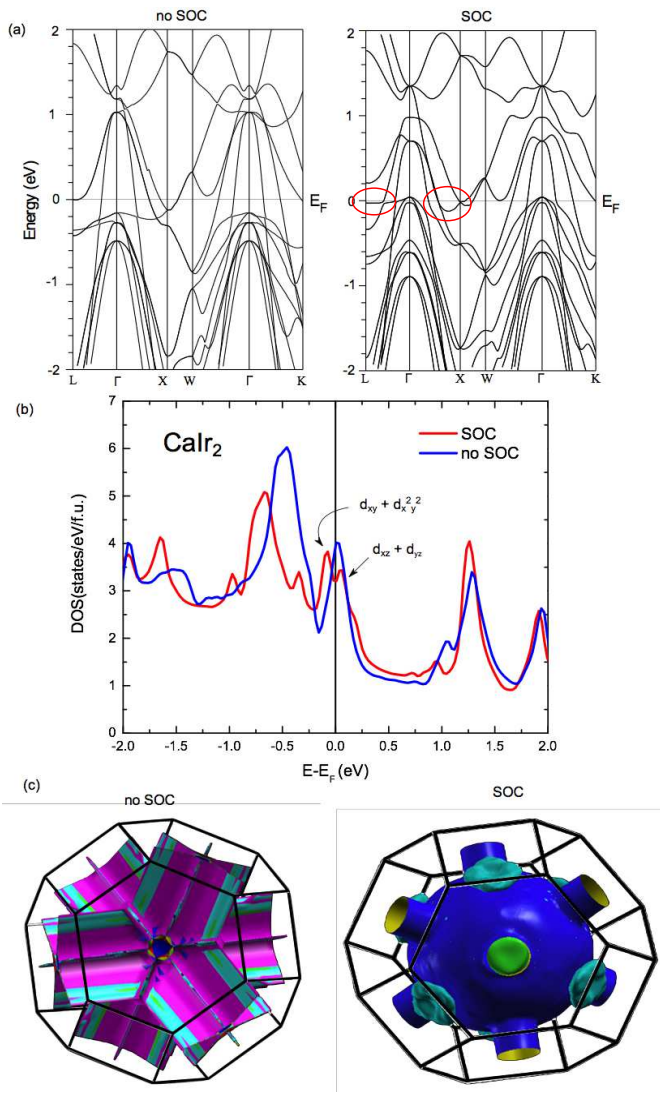}}
  \caption
    {
      (Color online) Analysis of the electronic structure of CaIr$_{2}$. (a) The electronic band structure,(b) The total DOS, and (c) the Fermi surface. The left sides of (a) and (c) show the behavior without spin orbit coupling and the right sides show the behavior with spin-orbit coupling.     
    }
  \label{Fig4}
\end{figure}

The ratio $\frac{\Delta C}{\gamma T_{c}}$ can be used to measure the strength of the electron-phonon coupling.~\cite{padamsee1973} The specific heat jump $\frac{\Delta C}{T_{c}}$ for the sample is about 8 $\frac{mJ}{mol~K^{2}}$, setting the value of $\frac{\Delta C}{\gamma~T_{c}}$ to 0.89. This is smaller than the weak-coupling limit of 1.43 for a conventional BCS superconductor. The results suggest that CaIr$_{2}$ is a weakly electron$-$phonon coupled superconductor.


In a simple Debye model for the phonon contribution to the specific heat, the $\beta$ coefficient is related to the Debye temperature $\Theta _{D}$ through $\beta = nN_{A}\frac{12}{5}\pi ^{4}R\Theta _{D}^{-3}$, where $R = 8.314~\frac{J}{mol~K}$, $\textit{n}$ is the number of atoms per formula unit and $N_{A}$ is Avogadro\textquoteright s number. The calculated Debye temperature for CaIr$_{2}$ is thus 160 K. This value of the Debye temperature is comparable to that of CeRu$_{2}$ (see Table.~\ref{tab:1}).
An estimation of the strength of the electron-phonon coupling can be derived from the McMillan formula
$\lambda _{ep} = \frac{1.04 + \mu ^{*} ln\frac{\Theta _{D}}{1.45T_{c}}}{(1-0.62\mu ^{*}) ln\frac{\Theta _{D}}{1.45T_{c}}-1.04}$.
McMillan\textquoteright s model contains the dimensionless electron-phonon coupling constant $\lambda _{ep}$, which, in the Eliashberg theory, is related to the phonon spectrum and the density of states.~\cite{mcmillan1968, poole1999}  This parameter $\lambda _{ep}$ represents the attractive interaction, while the second parameter $\mu ^{*}$ accounts for the screened Coulomb repulsion.
Using the Debye temperature $\Theta _{D}$, critical temperature $T_{c}$, and making the common assumption that $\mu ^{*} = 0.15$,~\cite{mcmillan1968} the electron-phonon coupling constant ($\lambda _{ep}$) obtained for CaIr$_{2}$ is 0.79, which suggests weak electron-phonon coupling behavior and agrees well with $\frac{\Delta C}{\gamma T_{c}} = 0.89 $.

\begin{table}[t]
\caption{Superconducting Parameters of the cubic Laves phase CaIr$_{2}$. Comparison is done with the cubic Laves phase CeRu$_{2}$. The superconducting parameters of CeRu$_{2}$ are extracted from Ref. 10,11.}
  \centering  
  \begin{tabular}{ lc   c   c   c }
  \hline \hline 
    Parameter & Units & CaIr$_{2}$ & CeRu$_{2}$    \\ \hline  \hline  
    $T_{c}$ & K & 5.8  &  6.2    \\
    $\rho _{0}$ & $m\Omega cm$ &  7.4  & 0.001    \\
    $\frac{dH_{c2}}{dT}\vert _{T=T_{c}}$ & $T~K^{-1}$ & -0.81  &  \\
    $\mu _{0}H_{c1}(0)$ & Oe & 381  & 200-400    \\
    $\mu _{0}H_{c2}(0)$ & T & 4.0  &  7.4    \\
    $\mu _{0}H^{Pauli}$ & T & 10.7  & 11.4   \\
    $\mu _{0}H (0)$ & T & 0.06  &  0.073   \\
    $\xi (0)$ & \AA & 90.7  &  67    \\
    $\lambda (0)$ & \AA & 960  & 1100-1400     \\
    $\kappa (0)$ & \AA & 10.6  & 19.4   \\
    $\gamma (0)$ & $\frac{mJ}{mol~K^{2}}$ & 8.4  &  29    \\
    $\frac{\Delta C}{\gamma T_{c}}$ & & 0.89  & 2.0    \\
    $\Theta _{D}$ & K & 160  & 173     \\
    $\lambda _{ep}$ &  & 0.79  &     \\
    $N(E_{F})$ & $\frac{eV}{f.u.}$ & 3.49  &      \\ \hline \hline
  \end{tabular}
  \label{tab:1}
\end{table}



The value of $\gamma$ extracted from the measured specific heat data corresponds to an electronic density of states at the Fermi energy $N(E_{F})(exp)$ of 0.70 states/(eV f.u.), as estimated from the relation~\cite{kittel, poole1999} $3\gamma = \pi ^{\frac{3}{2}} k_{B}^{2} N(E_{F}) (1+\lambda _{ep})$. This value is slightly lower than the value obtained from our band structure calculation. 
The actual H$_{c2}$ of real materials is generally influenced by both orbital and spin-paramagnetic
effects. The relative importance of the orbital and spin-paramagnetic effects can be described by the
Maki parameter,~\cite{maki}
which can be calculated as $\frac{\mu_{0}H_{c2}}{H^{Pauli}}=\frac{\alpha}{1.41}=0.52$.~\cite{carbotte, cdmgb22002, junod}  
The small Maki parameter obtained from this approximation is an indication that Pauli limiting is negligible. In contrast, a sizable Maki parameter is observed for CeRu$_{2}$, which suggests that a substantial spin-orbit component is present. Similar behavior was observed in some other unconventional superconductors such as Nb$_{0.18}$Re$_{0.82}$.~\cite{amar2011} 
The superconducting parameters of CaIr$_{2}$ are presented in Table.~\ref{tab:1}. Comparison with CeRu$_{2}$ is given because it shares the cubic Laves phase type structure with CaIr$_{2}$. 
The superconducting parameters of both compounds are comparable except that the CeRu$_{2}$ is a strongly electron-phonon coupled superconductor. 


Fig.~\ref{Fig4} shows the analysis of the band structure of CaIr$_{2}$ based on the ab-initio calculations. 
According to the calculated band structure, CaIr$_{2}$ is a three-dimensional metal; several bands with large dispersion cross the Fermi level (Fig.~\ref{Fig4}(a)). 
There are a couple of new features observed in the band structure (see circled areas of Fig. 4(a)) due to spin orbit coupling (SOC). A saddle point (at E$_{F}$) is visible along L- $\Gamma$ when SOC is present and as is band splitting along  $\Gamma$ - W. Saddle points in electronic band structures are instabilities that are known to cause materials to become magnetically ordered or superconducting.~\cite{irpyro, vhs} 
The bands at the Fermi level are all derived from Ir \textit{5d}-orbitals. Thus the pyrochlore arrangement of the Ir lattice, which determines the energy and dispersion of the bands from the Ir, has a significant effect on the electrons that become superconducting in CaIr$_{2}$. Similar dispersive bands from the pyrochlore lattice are observed for the 4\textit{d}-based cubic laves phase superconductor BaRh$_{2}$.~\cite{barh21991} 
The partial DOS shows that the total DOS is dominated by the contributions from the Ir sublattice and that the contribution from the Ca atoms near the Fermi level is almost negligible.
The total DOS (see Fig.~\ref{Fig4}(b)) shows that the Fermi level is located near the edge of a local maximum. Without spin orbit coupling (SOC) this peak is composed of degenerate Ir \textit{5d$_{xy}$+d$_{x^2y^2}$} and Ir \textit{5d$_{xz}$+d$_{yz}$} orbitals. SOC splits the degeneracy; the Ir \textit{5d$_{xy}$+d$_{x^2y^2}$} states move slightly below the Fermi level and the Ir \textit{5d$_{xz}$+d$_{yz}$} states move slightly above. Similar splitting is observed in oxide-based Ir pyrochlores\cite{irpyro}, however in that case the Ir \textit{5d} electrons are more localized resulting in the opening of a band gap. The magnitude of the splitting is roughly 0.3 eV in both cases. The value of the DOS at E$_{F}$ (SOC included) is comparable with the value estimated from the heat capacity data. Many bands are found in CaIr$_{2}$ through the Fermi energy, and therefore the calculated Fermi surface is complex. The Fermi surface is very strongly affected by the presence of the spin orbit coupling inherent to the 5\textit{d} element Ir, as seen in Fig.~\ref{Fig4}(c). 
Given the radical change in the Fermi surface due to the spin orbit coupling compared to the hypothetical case where no spin orbit coupling is present, one can speculate that CaIr$_{2}$ has T$_{c}$ on the high side of the Ir-based compounds due to the presence of the SOC. 

When the Laves phase structure contains \textit{f}-electrons then it seems to influence the electron-phonon coupling. To understand this picture we have compared the electronic band structure of the strongly coupled superconductor CeRu$_{2}$,~\cite{ceru2dft, celaru2dft} with the CaIr$_{2}$ electronic band structure. It can clearly be seen that the electronic structure of CeRu$_{2}$ is quite different from the CaIr$_{2}$ band structure. In CeRu$_{2}$, there are empty Ce-\textit{f}-states just above the Fermi level, while some of the \textit{f}-states are dispersed and hybridize mainly with Ru \textit{d}-states. In contrast, there is not much Ca-Ir hybridization observed in the \textit{s-d}-bands in CaIr$_{2}$. 
Considering the band structure and the derived parameters for CeRu$_{2}$, it appears that one of the major effects of the hybridization of Ce \textit{f-d} and Ce \textit{f} with Ru \textit{d} bands is to enhance the electron-phonon coupling. 
This shows Laves phases are favorable for superconductivity from an electron-phonon perspective; however, in addition, the electronic band structures play an important role.

In oxide based Ir pyrochlores, the pyrochlore lattice causes magnetic frustration and complex behavior.~\cite{irpyro} In the case of intermetallic CaIr$_2$ the bands are not localized and no magnetism is observed. The compound becomes superconducting, however, which might be linked to the special geometry of the pyrochlore lattice. Superconductivity can be often found close to instabilities, as for example van Hove singularities in the band structure.\cite{vhs} In our case we see a peak in the Density of States at the Fermi level which indicates the proximity of an instability. This could indicate that the pyrochlore lattice might be advantageous for finding superconductors.

\section{Conclusion}
We have characterized the properties of the pyrochlore lattice superconductor CaIr$_{2}$. The presence of a bulk superconducting transition with $T_{c}=5.8$~K was confirmed, and the properties of the superconductor were elucidated through magnetization and heat capacity measurements. The inferred electron-phonon coupling constant $\lambda _{ep}$ and $\frac{\Delta C}{\gamma T_{c}}$ show that CaIr$_{2}$ is a weakly coupled BCS-type superconductor. The electronic band structure calculations indicate that the the Ir \textit{d} states are dominant through the Fermi level. Given the profound effect of spin-orbit coupling on the calculated electronic structure, it can be argued that the value of $T_{c}$, and possibly even the existence of superconductivity at all, is due to the heavy element character imparted to this material by the Ir pyrochlore lattice.


\section{acknowledgments}
This work was supported by the Department of Energy, Division of Basic Energy Sciences, grant DE-FG02-98ER45706.

\end{document}